# Temporally Adjustable Longitudinal Fluid-Attenuated Inversion Recovery MRI Estimation / Synthesis for Multiple Sclerosis

Jueqi Wang, Derek Berger, Erin Mazerolle, Othman Soufan, Jacob Levman

Department of Computer Science, St Francis Xavier University, Canada
{x2019cwn, dberger, emazerol, osoufan, jlevman}@stfx.ca

**Abstract.** Multiple Sclerosis (MS) is a chronic progressive neurological disease characterized by the development of lesions in the white matter of the brain. $T_2$-fluid-attenuated inversion recovery (FLAIR) brain magnetic resonance imaging (MRI) provides superior visualization and characterization of MS lesions, relative to other MRI modalities. Longitudinal brain FLAIR MRI in MS, involving repetitively imaging a patient over time, provides helpful information for clinicians towards monitoring disease progression. Predicting future whole brain MRI examinations with variable time lag has only been attempted in limited applications, such as healthy aging and structural degeneration in Alzheimer's Disease. In this article, we present novel modifications to deep learning architectures for MS FLAIR image synthesis / estimation, in order to support prediction of longitudinal images in a flexible continuous way. This is achieved with learned transposed convolutions, which support modelling time as a spatially distributed array with variable temporal properties at different spatial locations. Thus, this approach can theoretically model spatially-specific time-dependent brain development, supporting the modelling of more rapid growth at appropriate physical locations, such as the site of an MS brain lesion. This approach also supports the clinician user to define how far into the future a predicted examination should target. Accurate prediction of future rounds of imaging can inform clinicians of potentially poor patient outcomes, which may be able to contribute to earlier treatment and better prognoses. Four distinct deep learning architectures have been developed. The ISBI2015 longitudinal MS dataset was used to validate and compare our proposed approaches. Results demonstrate that a modified ACGAN achieves the best performance and reduces variability in model accuracy. Public domain code is made available at https://github.com/stfxecutables/Temporally-Adjustable-Longitudinal-MRI-Synthesis.

# 1    Introduction

Multiple sclerosis (MS) is a chronic progressive neurological disease with a variable course [1] and has become a major cause of disability among young adults [2]. MS patients develop lesions in the white matter (WM) of the brain. Medical imaging plays an essential role as a diagnostic tool, where magnetic resonance imaging (MRI) is widely used for diagnosing MS because structural MRI can be used to image white matter (WM) lesions [3], and $T_2$-fluid-attenuated inversion recovery (FLAIR) MRI typically provides superior assessment of WM lesions than other commonly acquired sequences [2]. The development of WM lesions on follow-up MRI can be used to monitor disease progression and towards informing clinicians' treatment plans for MS patients [1]. Accurate prediction of future rounds of imaging in MS can warn clinicians as to unhealthy growth trajectories of patients with MS. Since prognoses are generally improved the earlier on in which the treatment begins, image prediction techniques have the potential to warn clinicians as to potential MS progression, and so, once highly accurate image prediction techniques are developed, they can inform clinicians and potentially form a critical component towards early treatment and improvement of clinical outcomes. Therefore, predicting future FLAIR MRI examinations could provide helpful information for clinicians in charge of managing patient care.

Recent studies have shown that deep generative models have the ability to predict future brain degeneration using MRI [4–7], [22-24]. Wegmayr et al. [4] proposed to use the Wasserstein-GAN model to generate synthetically aged brain images given a baseline scan. Their method needed to be applied recursively in order to predict different future time points, and could only predict into the future by multiples of a predefined time interval. In contrast, our model only requires one prediction and is supported by a single time lag input variable that can predict at any user-defined future point in time. Ravi et al. [5] proposed a 4D deep learning model, which could generate several future 3D scans from different time points at once. However, this method needs several time points across many participants, and requires an expanded architecture to produce multiple time point outputs. Wang et al. [7] proposed using several previous scans to predict the neurological MRI examination of a patient with Alzheimer's Disease (AD) using a U-Net. However, their method could only predict images at a fixed point in the future (6 months). Some studies [22, 24] require longitudinal images from two timepoints to predict future MRI scans.

Similar to our study, Xia et al. [6] proposed a 2D conditional GAN (cGAN) method, which also employs user-defined time as an input parameter alongside the subject scans into both the generator and discriminator, and predicts future scans at the target time point. They use an ordinal encoding of age with a $100 \times 1$ vector, which can only represent time information at discrete time intervals (such as annually). This ordinal encoding was incorporated into their novel deep learning architecture with a small bottleneck layer, which many common convolution neural network (CNN) models do not normally contain. Alternatively, our method supports a more flexible interval for temporal prediction, by simply providing the normalized time lag value, encoded in days between exams, into the learner. In our proposed approach, time information is first expanded using transposed convolutions, which is concatenated with internal feature maps in any CNN layer. In real-world clinical practice, the time between longitudinal exams for a central nervous system disorder (e.g., MS) is quite variable, and MS lesions have the potential to develop actively. We will distinguish their methods from ours more clearly in the methods section. Thus, the developments outlined in this paper have the potential to help extend image estimation / synthesis technologies to real-world clinical use. Additionally, several approaches have been proposed to use existing scans to predict future MS lesion progression [8–11], where the output of these models is lesion feature information instead of whole images. More recently, Kumar et al. [25] proposed a cGAN to generate counterfactual images to aid data-driven biomarker discovery and validated their method in a longitudinal MS dataset.

Despite those methods having shown great performance, most are concerned with predicting the healthy aging brain, as well as predicting AD MRI examinations. Predicting future brain FLAIR MRI examinations for MS patients is a topic that has not yet been fully explored. Thus, we are proposing deep learning models that can predict FLAIR images for MS patients at any user-defined amount of time into the future, while modelling time as a spatially distributed feature map, which allows for variable growth rate trajectories across different tissues, notably for brain lesions, which often progress / develop at different rates from healthy parenchyma. Our method could also be used as a novel data augmentation method for generating new samples for training deep neural networks.

Our work has four main contributions. First, we modify existing deep learning architectures with transposed convolutions to parameterize the time lag to prediction, which governs how far into the future to predict the next image. Second, the transposed convolution supports the modelling of time as a spatially distributed array of temporal variables, allowing the learning machine to model variable rates of growth distributed across brain tissues. Thus, the approach presented herein can support clinicians to estimate a patient's disease progression at multiple points in the future, and can model spatially variable tissue growth, atrophy and remission. The architecture modifications presented in this paper support the use of real-world longitudinal data whereby the time between scans is variable. Third, we developed modifications to 4 different deep learning architectures to add user-defined time lag using transposed convolutions: a modified U-Net, a generator-induced time GAN (gt-GAN), a discriminator-induced time GAN (dt-GAN) and a modified auxiliary classifier GAN (ACGAN). Fourth, we add an auxiliary classifier [15] in the discriminator in order to produce a performance improvement when compared with providing time lag information into both the discriminator and generator, as in a previous study [6].

## 2    Materials and Methods

### 2.1    Modeling time information by transposed convolution

This section illustrates our approach to providing time information into a CNN by transposed convolutions in order to predict future brain changes continuously.

We use transposed convolutions, instead of one-hot vectors [12] or ordinal binary vectors [6] used in previous studies, to expand the user selected time lag prediction variable to the same size as the input images, which theoretically supports the modelling of spatially-specific time-dependent brain development. Then we concatenate the learned spatially distributed feature map with the first layer of feature maps in the 3D U-Nets. We normalize the time information by using days between studies divided by 365, creating a floating point decimal number in years, which is more consistent with the nature of time as a continuous variable. Note that in [6], they also did an ablation study comparing normalized time information as one continuous variable (between 0 and 1) with their ordinal binary vector approach, which resulted in a network that would generate similar images to one another. In contrast, in our method, we first expand the time information by transposed convolution and then concatenate the result with our standard feature maps, while their method concatenated the continuous value with the image embedding directly. Our method also has the potential to flexibly add time information into any CNN model, while in [6], their ordinal binary vectors cannot be applied to every CNN model.

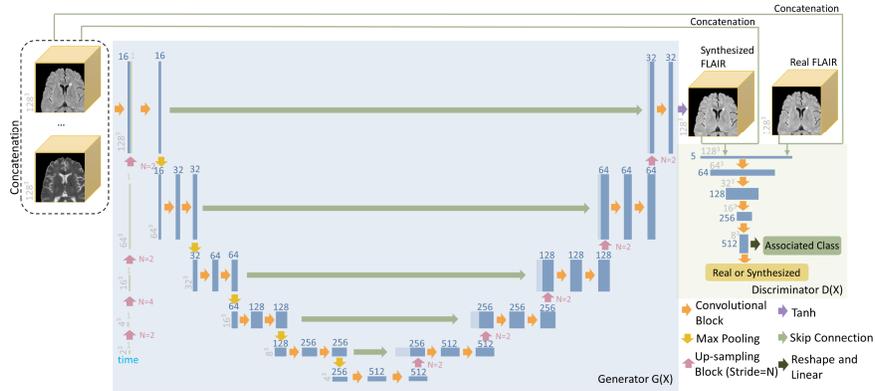

**Fig. 1.** Detailed architecture for modified ACGAN. Batch normalization and LeakyReLU with slope of 0.2 are used in the convolution block. Batch normalization was not used in the first blocks and last blocks of the Discriminator (D). The last convolution blocks in D used a sigmoid activation function.

### 2.2 Proposed architectures

Generative adversarial networks (GANs) are widely used for image synthesis. Conditional GANs (cGANs) [13] are more suitable for image-to-image translation problems by learning the condition of the input images. To this aim, we explored 4 different architectures for this application.

**3D U-Net.** As shown in Fig. 1, a 6-level 3D U-Net is utilized for all methods, acting as a baseline comparative model as well as the generator in the three subsequently developed GANs. We use the L1 distance as the loss function. The expanded time information is concatenated with feature maps after the first convolution block.

**Generator-Induced Time GAN (gt-GAN).** The second approach combined a discriminator and the L1 distance function for better perceptual performance and less blurring. The objective of the generator $G$ can be expressed as:

$$L^G_{gt-GAN} = E_{x,t}\left[\log\left(1 - D(x, G(x,t))\right)\right] + \lambda_{l1} E_{x,y,t}\left[||y - G(x,t)||_1\right] \quad (1)$$

where $E$ is the maximum likelihood estimation, $x, y$ denotes the input images and $x$'s corresponding target image after time $t$, separately. $\lambda_{l1}$ is a non-negative trade-off parameter, which is used to balance the adversarial and L1 loss. By minimizing this objective function, the generated image will not only fool the discriminator but also approximate the ground truth output at the pixel level [13]. As in Pix2pix [13], cGANs are trained by conditioning the learning model on the source image data. The discriminator $D$ takes both the source images and either a real target image or a synthesized one as input, and is tasked with predicting whether the image is real or not. The discriminator D is trained to maximize the following objective:

$$L^D_{gt-GAN} = E_{x,y}[\log D(x,y)] + E_{x,t}\left[\log\left(1 - D(x, G(x,t))\right)\right] \quad (2)$$

**Discriminator-Induced Time GAN (dt-GAN).** In dt-GAN, the time lag parameter is incorporated into both the generator and the discriminator $D$ using transposed convolutions. In this way, the time lag information is also learned by the discriminator, which could possibly help the discriminator to distinguish between the real images $y$ and synthesized images $G(x)$, towards potentially improving the generator's performance. The objective of the generator $G$ in dt-GAN is the same as gt-GAN. The objective of the discriminator D in dt-GAN is as follows:

$$L^D_{gt-GAN} = E_{x,y,t}[\log D(x,y,t)] + E_{x,t}[\log(1 - D(x, G(x,t), t))] \quad (3)$$

**Modified Auxiliary Classifier GAN (ACGAN).** Instead of providing the time information directly into the discriminator, the discriminator can potentially learn how to distinguish the difference between different time lags itself. Thus, the discriminator would learn to identify differences (i.e., the size of lesion areas) between different time lags to force the generator to generate better images. Based on this hypothesis, in the fourth approach, we used a modified auxiliary classifier GAN (ACGAN) [15]. For each given sample, there is input image $x$ and target image $y$, associated with the time lag $t$. In addition to that, we also classify each sample into a class label $c$ based on having similar time $t$. We add an auxiliary classifier on discriminator $D$. Fig. 1 demonstrates our proposed modified ACGAN architecture. The objective of the discriminator $D$ is:

$$L^D_{ACGAN} = E_{x,y}[\log D(x,y)] + E_{x,t}\left[\log\left(1 - D(x, G(x,t))\right)\right]$$
$$+ E_{x,y,c}[\log p(c|x,y)] + E_{x,c,t}[\log p(c|x, G(x,t))]$$

(4)

where $c$ is an associated label with each sample classified, based on the time lag input parameter. By maximizing this objective function, $D$ learns not only to distinguish whether this sample is a real one or not, but also to classify each sample into its corresponding class $c$. Simultaneously, $G$ tries to generate images that can be classified into the target class $c$, to enhance the accuracy of image synthesis [21].

### 2.3 Dataset and Evaluation Metrics

To validate our method, we used the ISBI2015 longitudinal MS dataset [16], which consists of 19 participants. Among them, 14 participants had scans at four time points, 4 participants had scans at five time points, and one had scans at six time points. All were acquired on the same MRI scanner. The first time-point MPRAGE was rigidly registered into 1 mm isotropic MNI template space and used as a baseline for the remaining images from the same time-point, as well as from each of the follow-up time-points. Consecutive time-points are separated by approximately one year for all participants in this dataset. The following modalities are provided for each time point: T1-w MPRAGE, T2-w, PD-w, and FLAIR. Our models predict images at varying time lags into the future, as such 139 samples are available in this dataset at varying time intervals. For instance, there are 6 samples from one participant with 4 time points (1→2, 1→3, 1→4, 2→3, 2→4, 3→4). All modalities from the early time-point and the user-defined time lag parameter were included to predict future FLAIR scans.

Three popular metrics are used in this study: peak signal-to-noise ratio (PSNR), normalized mean squared error (NMSE), and the structural similarity index (SSIM) [17].

### 2.4 Implementation details

We cropped out an image size of (150, 190, 150) to reduce the background region. Each volume was linearly scaled to [-1, 1] from the original intensity values for normalization. To fit the 3D image into the generator and make the whole model fit into GPU memory, we split them into eight overlapping patches of size (128, 128, 128). The overlapped regions are averaged to aggregate those patches. A data augmentation of rotation with random angle $[-12°, 12°]$ and a random spatial scaling factor $[0.9, 1.1]$ was employed during training. Batch size was 3 for all methods. 5-fold cross validation was applied at the participant level to effectively evaluate different methods (2 folds have 4 4-time-point participants; one fold has 3 4-time-point and one 5-time-point participants; one fold has 2 4-time-point and one 5-time-point participant with the last fold having one 4-time-point, 2 5-time-point and 1 6-time-point participants). Samples are grouped into different classes $c$ based on rounding off the time between the input exams and the target predicted exams to a whole year value in modified ACGAN. We use the Adam optimizer [18] with momentum parameters $\beta_1 = 0.5$ and $\beta_2 = 0.999$ and weight decay $\lambda = 7 \times 10^{-8}$ to optimize all the networks.

PatchGAN [13] was used for penalizing each patch to be real or fake to support the discriminator in the GAN to encourage high quality image generation. As in [19], $\lambda_{l1}$ was set to 300 for all cGANs during training. To balance the generator and discriminator in GANs, we use label smoothing [20] to improve the stability of training GANs. The learning rate was set to 0.0002 for both the generator and discriminator in all the GANs during the first 150 epochs, then, linearly decaying to 0 for the following 50 epochs. For the baseline modified U-Net, the learning rate was set to $7 \times 10^{-5}$ for the first 150 epochs, then linearly decaying to 0 for the following 50 epochs. Experiments were performed on 4 Nvidia A100 GPUs with 40 GBs of RAM using distributed data in parallel via the PyTorch framework. Training took around 5 hours for each fold for each model.

## 3      Results and Discussion

Table 1 shows the quantitative results obtained by different methods that we investigated in terms of mean PSNR, NMSE, and SSIM values and their corresponding standard deviation. All the metrics are computed on the aggregated 3D volume instead of patches to represent the performance on the whole scans. We linearly scaled each volume to [0, 1] before computing all the metrics to ensure a fair comparison. First, we observe that all the GANs provide better results than the baseline modified U-Net. Nevertheless, by integrating the time lag parameter into both the generator $G$ and the discriminator $D$, dt-GAN does not achieve better performance in all the three metrics as compared with gt-GAN, which only integrates the time lag parameter $t$ into the generator $G$. This might confirm that integration of the time lag into both the generator and the discriminator cannot improve image synthesis performance in this situation. The modified ACGAN achieves the best results and the smallest standard error across all three performance metrics.

**Table 1.** Quantitative Evaluation Results of Different Methods (mean ± standard deviation), obtained by evaluated methods on the validation folds.[1]

| Methods | PSNR ↑ | NMSE↓ | SSIM ↑ |
|---|---|---|---|
| Modified ACGAN | **28.8721±2.709** | **0.2006±0.080** | **0.9148±0.024** |
| dt-GAN | 27.4969±2.851 | 0.2368±0.095 | 0.9068±0.026 |
| gt-GAN | 28.4069±3.136 | 0.2160±0.099 | 0.9089±0.027 |
| Modified U-Net | 22.9473±3.655 | 0.4296±0.195 | 0.8931±0.031 |

Qualitative results of the proposed modified ACGAN are illustrated in Fig. 2. With respect to participant A, the source image's expanded region-of-interest (ROI) exhibits three subtle lesions that are changing temporally between the source and target acquisitions, which in this examination were 3 years apart. Note that both lesions marked by a red circle appear to have gone into remission and are extremely difficult to visually identify on the target image. Also noteworthy is that the subtle lesion on the source exam, marked by a red arrow, developed into a more prominent lesion by the target image acquisition. Our proposed modified ACGAN approach to image prediction has resulted in a reduction of visual lesion prominence for both lesions exhibiting remission (marked by red circles), as well as increased visual prominence for the expanding lesion marked by a red arrow. With respect to participant B, the red circled lesion exhibits a hypointensity on the target image which likely implies the development of regional atrophy not present in the original source image. Our modified ACGAN approach was able to partially model this hypointensity's developmental trajectory, potentially reflective of tissue atrophy. These results from both participants imply that our proposed approach is capable of modeling subtle lesion growth, lesion remission, as well as a limited amount of lesion atrophy.

---

[1]   We cannot report the metrics only based on the lesion area, since no lesion labels were provided to 14 participants in this ISBI2015 dataset.

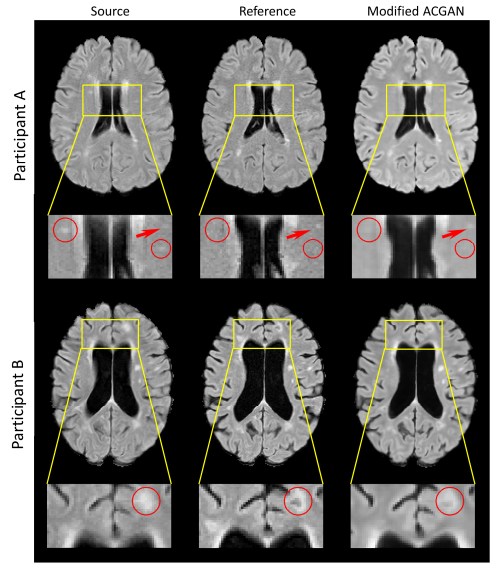

**Fig. 2.** Example predicted images in the validation fold from the leading modified ACGAN. Participant A: Source is FLAIR scan from time point 1, Target is FLAIR from time point 4, ACGAN is predicting images using modalities from time point 1 and the user-defined time lag parameter was set to predict time point 4 FLAIR. Participant B: Source is time point 1 FLAIR, Target is time point 5 FLAIR, ACGAN is predicting images using modalities from time point 1 and the user-defined time lag parameter was set to predict time point 5 FLAIR.

Future work could involve the use of ROI specific weighted loss, in order to increase the ability of the network to focus on small lesion areas. Although Fig. 2 demonstrates small changes to existing lesions or the appearance of new lesions, our overall loss function is expected to be dominated by whole brain factors. Thus, a model with ROI specific weighted loss could be more valuable for clinical interest, and combining ROI specific loss with the approach presented herein is the subject of future work. One of the limitations of the leading modified ACGAN is an associated class needed to be assigned to each sample, for which it is difficult to find a priority solution, when there is large variability between time intervals. A potential solution to this problem is to use clustering algorithms (i.e., K-nearest neighbors algorithm, etc.) to define the respective class labels. Future work will also examine additional datasets with more variability in the time between examinations, as this dataset largely consists of examinations acquired at yearly intervals. This dataset did not include gold-standard ROIs for most of the MS lesions, as such, we were unable to report lesion specific performance metrics in Table 1. Future work will involve evaluating the proposed approach on datasets with provided gold-standard ROIs, as well as evaluating the proposed approach on datasets such as the one we have used in this study with additional segmentation technology to automatically define the lesion ROIs, to assist in evaluating lesion-specific predictive performance.

## 4 Conclusion

In this work, we propose a new way to integrate time lag information into deep learning models by transposed convolutions, to predict future brain FLAIR MRI examinations for MS patients. We also compared 4 different approaches to provide time lag input parameters into cGANs and the U-Net. By using transposed convolutions, the time lag information has the potential to be spatially distributed and concatenated into any CNN architecture layer. Our method could also create a more flexible interval in a continuous way, which is more suitable for MS to help extend image estimation / synthesis technologies to real-world clinical use. We also propose to use an auxiliary classifier in the discriminator, which has potential to boost predictive accuracy. A longitudinal MS dataset with larger participant size and more timepoints with each participant will be more valuable for validating of this method.

## Acknowledgements

This work was supported by an NSERC Discovery Grant to JL. Funding was also provided by a Nova Scotia Graduate Scholarship and a StFX Graduate Scholarship to JW. Computational resources were provided by Compute Canada.

## References

1. McGinley, M.P., Goldschmidt, C.H., Rae-Grant, A.D.: Diagnosis and Treatment of Multiple Sclerosis: A Review. JAMA. 325, 765–779 (2021).
2. Wei, W., Poirion, E., Bodini, B., Durrleman, S., Colliot, O., Stankoff, B., Ayache, N.: Fluid-attenuated inversion recovery MRI synthesis from multisequence MRI using three-dimensional fully convolutional networks for multiple sclerosis. J Med Imaging (Bellingham). 6, 14005 (2019). https://doi.org/10.1117/1.JMI.6.1.014005.

3. Salem, M., Valverde, S., Cabezas, M., Pareto, D., Oliver, A., Salvi, J., Rovira, A., Llado, X.: Multiple Sclerosis Lesion Synthesis in MRI Using an Encoder-Decoder U-NET. IEEE Access. 7, 25171–25184 (2019). https://doi.org/10.1109/ACCESS.2019.2900198.
4. Wegmayr, V., Hörold, M., Buhmann, J.M.: Generative Aging Of Brain MRI For Early Prediction Of MCI-AD Conversion. In: 2019 IEEE 16th International Symposium on Biomedical Imaging (ISBI 2019). pp. 1042–1046 (2019). https://doi.org/10.1109/ISBI.2019.8759394.
5. Ravi, D., Blumberg, S.B., Ingala, S., Barkhof, F., Alexander, D.C., Oxtoby, N.P.: Degenerative adversarial neuroimage nets for brain scan simulations: Application in ageing and dementia. Medical Image Analysis. 75, 102257 (2022). https://doi.org/https://doi.org/10.1016/j.media.2021.102257.
6. Xia, T., Chartsias, A., Wang, C., Tsaftaris, S.A.: Learning to synthesise the ageing brain without longitudinal data. Medical Image Analysis. 73, 102169 (2021). https://doi.org/10.1016/J.MEDIA.2021.102169.
7. Wang, J., Berger, D., Mattie, D., Levman, J.: Multichannel input pixelwise regression 3D U-Nets for medical image estimation with 3 applications in brain MRI. In: International conference on Medical Imaging with Deep Learning (2021).
8. Doyle Andrew and Precup, D. and A.D.L. and A.T.: Predicting Future Disease Activity and Treatment Responders for Multiple Sclerosis Patients Using a Bag-of-Lesions Brain Representation. Medical Image Computing and Computer Assisted Intervention − MICCAI 2017. pp. 186–194. Springer International Publishing, Cham (2017).
9. Tousignant, A., Lemaître, P., Precup, D., Arnold, D.L., Arbel, T.: Prediction of Disease Progression in Multiple Sclerosis Patients using Deep Learning Analysis of MRI Data. In: Cardoso, M.J., Feragen, A., Glocker, B., Konukoglu, E., Oguz, I., Unal, G., and Vercauteren, T. (eds.) Proceedings of The 2nd International Conference on Medical Imaging with Deep Learning. pp. 483–492. PMLR (2019).
10. Sepahvand Nazanin Mohammadi and Hassner, T. and A.D.L. and A.T.: CNN Prediction of Future Disease Activity for Multiple Sclerosis Patients from Baseline MRI and Lesion Labels. Brainlesion: Glioma, Multiple Sclerosis, Stroke and Traumatic Brain Injuries. pp. 57–69. (2019).
11. Durso-Finley, J., Falet, J.-P.R., Nichyporuk, B., Arnold, D.L., Arbel, T.: Personalized Prediction of Future Lesion Activity and Treatment Effect in Multiple Sclerosis from Baseline MRI. International conference on Medical Imaging with Deep Learning. 1–20 (2022).
12. Zhang, Z., Song, Y., Qi, H.: Age progression/regression by conditional adversarial autoencoder. In: Proceedings of the IEEE conference on computer vision and pattern recognition. pp. 5810–5818 (2017).
13. Isola, P., Zhu, J.-Y., Zhou, T., Efros, A.A.: Image-to-Image Translation with Conditional Adversarial Networks. In: 2017 IEEE Conference on Computer Vision and Pattern Recognition (CVPR). pp. 5967–5976 (2017).
14. Goodfellow, I.J.: NIPS 2016 Tutorial: Generative Adversarial Networks. arXiv preprint arXiv:1701.00160. (2016).
15. Odena, A., Olah, C., Shlens, J.: Conditional image synthesis with auxiliary classifier GANs. In: ICML'17 Proceedings of the 34th International Conference on Machine Learning - Volume 70. pp. 2642–2651 (2017).
16. Carass, A., Roy, S., Jog, A., Cuzzocreo, J.L., Magrath, E., Gherman, A., Button, J., Nguyen, J., Prados, F., Sudre, C.H., Jorge Cardoso, M., Cawley, N., Ciccarelli, O., Wheeler-Kingshott, C.A.M., Ourselin, S., Catanese, L., Deshpande, H., Maurel, P., Commowick, O., Barillot, C., Tomas-Fernandez, X., Warfield, S.K., Vaidya, S., Chunduru, A., Muthuganapathy, R., Krishnamurthi, G., Jesson, A., Arbel, T., Maier, O., Handels, H., Iheme,

L.O., Unay, D., Jain, S., Sima, D.M., Smeets, D., Ghafoorian, M., Platel, B., Birenbaum, A., Greenspan, H., Bazin, P.-L., Calabresi, P.A., Crainiceanu, C.M., Ellingsen, L.M., Reich, D.S., Prince, J.L., Pham, D.L.: Longitudinal multiple sclerosis lesion segmentation: Resource and challenge. Neuroimage. 148, 77–102 (2017).
17. Wang, Z., Bovik, A.C., Sheikh, H.R., Simoncelli, E.P.: Image quality assessment: from error visibility to structural similarity. IEEE Transactions on Image Processing. 13, 600–612 (2004). https://doi.org/10.1109/TIP.2003.819861.
18. Kingma, D.P., Ba, J.L.: Adam: A Method for Stochastic Optimization. In: ICLR 2015 : International Conference on Learning Representations 2015 (2015).
19. Yu, B., Zhou, L., Wang, L., Shi, Y., Fripp, J., Bourgeat, P.: Ea-GANs: Edge-Aware Generative Adversarial Networks for Cross-Modality MR Image Synthesis. IEEE Transactions on Medical Imaging. 38, 1750–1762 (2019). https://doi.org/10.1109/TMI.2019.2895894.
20. Salimans, T., Goodfellow, I., Zaremba, W., Cheung, V., Radford, A., Chen, X.: Improved techniques for training GANs. In: NIPS'16 Proceedings of the 30th International Conference on Neural Information Processing Systems. pp. 2234–2242 (2016).
21. Choi, Y., Choi, M., Kim, M., Ha, J.-W., Kim, S., Choo, J.: Stargan: Unified generative adversarial networks for multi-domain image-to-image translation. In: Proceedings of the IEEE conference on computer vision and pattern recognition, pp. 8789-8797. (2018)
22. Fu, J., Tzortzakakis, A., Barroso, J., Westman, E., Ferreira, D., Moreno, R.: Generative Aging of Brain Images with Diffeomorphic Registration. arXiv preprint arXiv:2205.15607 (2022)
23. Bowles, C., Gunn, R., Hammers, A., Rueckert, D.: Modelling the progression of Alzheimer's disease in MRI using generative adversarial networks. SPIE (2018)
24. Kim, S.T., Küçükaslan, U., Navab, N.: Longitudinal Brain MR Image Modeling Using Personalized Memory for Alzheimer's Disease. IEEE Access 9, 143212-143221 (2021)
25. Kumar, A., Hu, A., Nichyporuk, B., Falet, J.-P.R., Arnold, D.L., Tsaftaris, S., Arbel, T.: Counterfactual Image Synthesis for Discovery of Personalized Predictive Image Markers. In: Medical Image Assisted BIomarkers' Discovery. (2022)